# Performance Cost Tradeoffs in Intelligent Load Balancing for Multi Data Center Cloud Systems: From Static Policies to Adaptive Resource Distribution

Saeid Aghasoleymani Najafabadi[1], Elaheh Nabavi Nia[2]

1-Faculty of Industrial Engineering, Urmia University of Technology, Urmia, Iran

2-Harrisburg University of Science and Technology, McLean, VA, USA

**Abstract**

Cloud computing infrastructures increasingly rely on geographically distributed data centers to meet the growing demand for low latency, high availability, and cost-efficient service delivery. In this context, load balancing plays a critical role in optimizing resource utilization while maintaining acceptable quality of service (QoS) under dynamic and heterogeneous workloads. This study presents a comprehensive performance and cost evaluation of three widely used load balancing strategies, namely Round Robin, Equally Spread Current Execution Load, and Throttled, within a multi data center cloud environment using the Cloud Analyst simulation framework. Multiple deployment scenarios are examined by varying data center locations, user base distribution, network latency, and workload intensity. Key performance metrics, including overall response time, data center processing time, request handling behavior, and operational cost such as virtual machine and data transfer costs, are analyzed across two strategy steps. The results indicate that while the Round Robin strategy achieves lower internal processing times, the Equally Spread and Throttled strategies provide improved workload stability and reduced peak response times under high demand conditions. Furthermore, distributing resources across multiple data centers significantly reduces user perceived latency and enhances system scalability, albeit with associated cost tradeoffs. The findings demonstrate that no single load balancing strategy is universally optimal; instead, performance and cost efficiency depend on workload characteristics, geographic distribution, and system objectives. This work offers practical insights for cloud service providers and system designers, emphasizing the importance of intelligent resource distribution and adaptive load balancing policies for sustainable and high-performance cloud infrastructures.

***Keywords:*** Cloud Computing Performance, Load Balancing Strategies, Resource Distribution Optimization, Cloud Analyst Simulation, Operational Cost Analysis.

## 1. Introduction

During the modern digital era, cloud computing has emerged as a crucial technological paradigm, revolutionizing the way computing resources are accessed, managed, and delivered. With its on-demand network access to configurable computing resources, cloud computing offers unparalleled scalability, flexibility, and cost-efficiency benefits. A pay-as-you-go billing model facilitates users' access to a wide range of services, from data storage to processing power. Cloud service providers must navigate many challenges to maintain and enhance quality-of-service delivery because of the rapid adoption and increased reliance on cloud services. In a cloud environment, load balancing is a critical process that involves the distribution of workloads and computing resources. A load balancer ensures that no single server or resource is overwhelmed by demand, thereby avoiding bottlenecks and maximizing resource utilization. In essence, load balancing optimizes cloud service performance, reduces execution costs, improves stability, and minimizes response times. As a result, cloud services provided to end users are improved significantly. This optimization is crucial as cloud platforms support an ever-expanding range of complex applications, from real-time augmented reality systems to advanced medical diagnostics. (Norcéide et al., 2024; Farhadi Nia et al., 2025).

Static and dynamic load balancing algorithms are at the heart of this process. In environments where workloads are evenly distributed and predictable, static load balancing algorithms work well based on predetermined information about the system's resources. In addition to round-robin algorithms, there are also First-Come-First Serve (FCFS) algorithms and threshold algorithms. As a result, these algorithms are not designed to account for the real-time state of the system, which can result in inefficiencies in rapidly changing or unpredictable cloud computing environments. A dynamic load balancing algorithm, on the other hand, makes decisions about workload distribution based on the current state of the system, including the availability and demand of resources. Dynamic algorithms are particularly useful in cloud computing scenarios characterized by highly variable and unpredictable demand. Dynamic load balancing algorithms improve cloud services' responsiveness and efficiency by continuously monitoring and adjusting resources in real-time. A dynamic approach to load balancing is exemplified by soft computing methodologies, such as genetic algorithms, ant colony optimization, and honeybee algorithms. To ensure equitable resource utilization and maintain system equilibrium, these algorithms leverage intuitive problem-solving techniques to find optimal or near-optimal solutions. Such advanced computational methods are not unique to cloud management and are being adapted for complex classification and segmentation tasks in other fields, like using improved Convolutional Neural Network (CNN) and U-Net models (Mahdavi, 2022a). It is impossible to overstate the importance of load balancing in cloud computing, as it directly impacts performance, reliability, and scalability. In addition to ensuring that resources are efficiently utilized, an effective load balancing strategy prevents individual servers from being overloaded, which could result in service degradation or outages. In addition, load balancing mechanisms reduce latency and improve response time of cloud services by optimizing the distribution of workloads. Furthermore, cloud computing has brought about a paradigm shift in organizational IT infrastructure (Ahmadi et al. 2023). Cloud services allow organizations, especially small and medium-sized businesses, to access high-quality computing resources without the need for significant upfront investments in hardware and infrastructure. Organizations have been able to scale their operations dynamically, adapt to changing market demands more quickly, and innovate more quickly because of this democratization of access to computing resources. Load balancing remains a critical component of leveraging the full potential of cloud computing. The continuous improvement of these underlying systems is essential for supporting specialized academic and research endeavors, such as the development of educational modules for complex biological systems. (Nia et al., 2023).

To address the challenges posed by the dynamic and distributed nature of cloud environments, advanced load balancing algorithms will play a crucial role in developing and implementing load balancing algorithms. Research and experimentation in this field are aimed at developing more sophisticated, adaptive, and efficient load balancing solutions that can meet the needs of modern cloud computing applications. To reinforce the foundational pillars of cloud computing, cloud service providers can continuously refine these algorithms to ensure that their infrastructures are resilient, flexible, and capable of delivering high-quality services to an ever-growing number of users.

## 2. Literature review

It has emerged as a critical concern in the realm of cloud computing to optimize resource utilization and improve system performance through efficient task scheduling and load balancing. Various approaches and methodologies are available to address these challenges, according to a review of recent literature. The BCSV scheduling algorithm is introduced by Chiang et al. (2023) within the context of cloud computing. Prior algorithms such as Suffrage, MaxSuffrage, and AMS overlooked load balancing issues. BCSV algorithm significantly enhanced task dispatch performance by incorporating Smallest Suffrage Value (SSV), Largest Suffrage Value (LSV), and Criteria Suffrage Value (CSV). Its superiority in achieving better load balance and makespan across heterogeneous network environments marks a notable advance in task scheduling and load balancing methods. Towards cost-efficiency and equitable job distribution in cloud computing, Sarhadi and Torkestani (2023) proposed a novel approach. By leveraging learning automata for reinforcement learning, their algorithms prevent disproportionate cost increases for individual machines while ensuring balanced utilization. Using the CloudSim toolkit, their approach was validated to be more efficient and balanced in terms of Final Cost and Total Cost, outperforming existing algorithms such as BCO, PES, CJS, PPO, and MCT. In this study, learning automata are explored as a potential tool for improving the economic and operational efficiency of cloud computing. They evaluated the performance of particle swarm optimization (PSO), round robin (RR), equally spread current execution (ESCE), and throttled load balancing algorithms. They meticulously compared these algorithms against various service broker policies using the Cloud Analyst platform. This study provided critical insights into the efficiency of these algorithms when applied to different workloads and user bases by highlighting optimized response times, data center processing times, and overall cost metrics. Comparing load-balancing strategies leads to a better understanding of their practical applications and limitations. Safari et al. (2025) examined the role of sentiment analysis in platform-based digital markets, showing that sentiment polarity extracted from user-generated content significantly influences user behavior and decision-making outcomes.

Tripathy et al. (2023) studied mist-fog cloud computing, which addresses the latency and power consumption challenges inherent in cloud computing's centralized datacenters. Based on their review, they proposed a layered architecture that integrates IoT, Mist, Fog, and Cloud layers to distribute loads uniformly and reduce latency. In addition to providing a comprehensive overview of current load balancing strategies, this study also identifies critical issues, challenges, and directions for future research. Ebneyousef and Shirmarz (2023) offered a detailed survey of load balancing algorithms and approaches in fog computing. Using literature from 2018 to 2022, they categorized algorithms, system architectures, tools, and applications, highlighting their advantages and disadvantages. Researchers and practitioners can use this taxonomy to improve fog computing load balancing performance, especially for delay-sensitive IoT applications. To form a hybrid PSO_PGA algorithm, Fu et al. (2023) combined particle swarm optimization and genetic algorithms. To enhance the search capabilities within the solution space, this method uses phagocytosis and genetic algorithm operations such as crossover and mutation. Compared to existing algorithms, their algorithm demonstrated significant improvements in cloud task completion times and convergence accuracy. In cloud environments, hybrid evolutionary algorithms have the potential to overcome the limitations of traditional task scheduling methods. A comparison of throttled load balancing with round robin and equally spread current execution (ESCE) algorithms was presented by Mohapatra et al. (2023). By optimizing resource utilization and satisfying user demands, load balancing algorithms are critical to improving cloud performance. Based on their evaluation of these established algorithms, the authors identified areas for further improvement in cloud computing load balancing practices. Safari et al. (2025) investigated the application of machine learning techniques for improving resilience and anomaly detection in distributed and resource-constrained systems, emphasizing the importance of adaptive models capable of handling evolving threat patterns. Their findings demonstrate that conventional classifiers often struggle to generalize beyond known behaviors, reinforcing the need for statistically grounded approaches to unknown attack detection.

Hodžić and Mrdović (2023) proposed a novel approach for balancing cloud workloads using genetic algorithms. The methodology relies on instantaneous processing of individual requests upon arrival, aiming to distribute user requests evenly among cloud resources. Test simulations indicated superior response and processing times over round robin, ESCE, and throttled algorithms, as well as an improvement over another genetic-based algorithm, DTGA. Genetic algorithms are effective in tackling the load balancing challenge, which presents a promising direction for future research. Santhosh and Vishnu (2023) conducted a case study on virtual machine load balancing algorithms in cloud data centers, emphasizing resource utilization and scheduling. Round robin, throttled, and ESCE load balancing techniques are discussed in their work. Studying load balancing algorithms and their impact on cloud computing environments contributes to a better understanding of their operational dynamics. Based on static and dynamic load balancing algorithms, Garg et al. (2023) optimized cloud service performance. To address the challenges of resource stalemate during allocation, they proposed a strategy that considers various response times and waiting times. Their approach successfully reduces response times and enhances the overall efficiency of cloud services, suggesting the potential benefits of integrating both static and dynamic load balancing strategies in cloud computing to mitigate overhead, starvation, and deadlock.

**Table 1:** Summary of recent literature on load balancing algorithms and applications in cloud computing

| Author(s) | Year | Load Balancing | Aim | Key Results |
|---|---|---|---|---|
| Haris, M., & Zubair, S. | 2025 | Hybrid Battle Royale Deep Reinforcement Learning (BRDRL) | To enhance the efficiency of load balancing by combining Deep Reinforcement Learning (DRL) and Battle Royale Optimization (BRO) to select the best under-loaded VM for task migration. | The proposed algorithm performed 3.9% better in throughput and 15.3% better in response time when compared to standard DRL methods. |
| Krishna Nayak, R., & Giduturi Srinivasa Rao | 2025 | SpiWasp-optimized Hybrid Convolution Network with Bidirectional Long Short-Term Memory (SpiWasp-CNTM) | To address workload imbalances and bottlenecks by using an optimal deep learning model that adeptly processes both spatial and temporal patterns in dynamic workloads. | The model demonstrated high reliability and superior performance, achieving low error metrics (MAE of 1.12, RMSE of 2.09), indicating its effectiveness in optimizing load distribution. |
| Sonia, & Nath, R. | 2025 | Systematic Review of ML/DL Approaches | To systematically review load balancing methods in cloud computing that use machine learning and deep learning, focusing on trends, challenges, and solutions. | The review highlights that machine and deep learning-based approaches (including reinforcement learning and neural networks) show great promise in adapting to changing workloads to optimize resource allocation and improve system performance. |
| Rashid, Y., & Nakpih, C. I. | 2024 | Response Time Efficient Load Balancer (ReT-ELBa) | To introduce and evaluate a novel algorithm that allocates tasks based on their size and VM state to minimize response time and increase resource utilization. | In simulations using Cloud Analyst, ReT-ELBa outperformed both Throttled and Round Robin algorithms in relation to response time and data center processing time |
| Syed, D., Muhammad, G., & Rizvi, S. | 2024 | Systematic Review of Metaheuristic Algorithms | To provide a detailed systematic review of load balancing techniques, identifying metaheuristic-based dynamic algorithms as an optimal solution for handling increased traffic. | The review confirms that metaheuristic algorithms are superior to static and traditional dynamic methods for cloud load balancing, and it provides a comparative analysis of their performance based on various QoS parameters |
| Chiang et al. | 2023 | BCSV Scheduling Algorithm | To improve task scheduling in cloud computing by addressing load balancing issues. | Achieved better load balance and make span than existing algorithms in various network environments. |
| Sarhadi, & Torkestani. | 2023 | Learning Automata-Based Algorithms | To enhance cost-efficiency and load balancing in cloud computing using learning automata. | Outperformed existing algorithms in terms of efficiency and balance between final and total cost. |
| Shahid et al. | 2023 | Evaluation of PSO, RR, ESCE, and Throttled Algorithms | To evaluate performance of load-balancing algorithms in cloud computing. | Provided detailed performance comparisons, highlighting efficiency metrics across different workloads. |
| Tripathy et al. | 2023 | Review of Load Balancing Algorithms for Mist-Fog-Cloud Paradigm | To review and suggest future directions for load balancing in mist-fog-cloud computing. | Proposed a layered architecture and identified critical issues and challenges in the paradigm. |
| Ebneyousef, & Shirmarz | 2023 | Taxonomy of Load Balancing Algorithms in Fog Computing | To survey and categorize load balancing algorithms in fog computing for IoT applications. | Offered insights into advantages and disadvantages of various algorithms, emphasizing performance improvement. |
| Fu et al. | 2023 | Hybrid Particle Swarm and Genetic Algorithm (PSO_PGA) | To optimize task scheduling in cloud computing through a hybrid algorithm. | Demonstrated improvements in overall completion time and convergence accuracy over existing methods. |
| Mohapatra et al. | 2023 | Enhanced Throttled Load Balancing Algorithm | To optimize cloud computing efficiency by improving performance through efficient load balancing. | Comprehensive analysis of round robin, equally spread current execution, and throttled algorithms, identifying strengths and weaknesses to suggest improvements. |
| Hodžić & Mrdović | 2023 | Genetic Algorithms for Load Balancing | To distribute user requests evenly across cloud resources with better response and processing times. | Outperformed round robin, ESCE, throttled, and DTGA genetic-based algorithm in response and processing times. |
| Santhosh & Vishnu | 2023 | Round Robin, Throttled, and Equally Spread Current Execution Load Balancing Techniques | To understand and clarify various scheduling techniques and service broker algorithms in cloud computing. | Provided an overview and clarified the operational dynamics of load balancing algorithms in cloud data centers. |
| Garg et al. | 2023 | Static and Dynamic Load Balancing Algorithms | To optimize cloud service performance by reducing response times and avoiding resource stalemates. | Demonstrated a reduction in response times and an increase in overall cloud efficiency, overcoming limitations such as overhead, starvation, and deadlock. |
| Kehoe et al. | 2015 | Cloud Robotics and Automation | To explore benefits of cloud infrastructure for robotics and automation systems. | Identified potential of cloud for big data access, computing, collective learning, and human computation in robotics. |
| Bera et al. | 2014 | Survey on Cloud Computing Applications for Smart Grid | To survey cloud computing applications in smart grid for energy, information management, and security. | Highlighted how cloud applications can enhance smart grid development, identifying research problems. |
| Shoja et al. | 2014 | Comparative Survey on Load Balancing Algorithms | To compare existing load balancing algorithms in cloud computing based on key metrics. | Discussed strengths and weaknesses of Round Robin and Throttled algorithms. |
| Supriya et al. | 2014 | Hierarchical Trust Model | To estimate trust values for cloud service providers based on varying levels of trustworthiness. | Compared trust values of various providers, enhancing consumer confidence in cloud services. |
| Supriya et al. | 2013 | Fuzzy Logic-Based Trust Management Model | To compare cloud service providers based on trust using a fuzzy logic model. | Proposed a model to objectively evaluate service levels for quality, reliability, and security. |

Using key metrics such as response time and data processing time, Shoja et al. Azizi (2014) compared existing load balancing algorithms in cloud computing. Round Robin and throttled scheduling algorithms were discussed, emphasizing the importance of efficient load balancing mechanisms in reducing energy consumption, maximizing resource utilization, and minimizing job rejections. As a result of this comparative analysis, future efforts to refine cloud computing performance will be guided by insights into strengths and weaknesses of different load balancing strategies. A study by Bera et a. (2014) examined the role of energy management, information management, and security within the smart grid. They identified how cloud computing can address the challenges of conventional smart grid systems and pointed to future opportunities for leveraging cloud technology to boost smart grid reliability, efficiency, security, and cost-effectiveness. By suggesting future directions for research and development, this review contributes to a broader understanding of the potential intersection of cloud computing and energy systems. This intersection is an active area of research, with studies focusing on intelligent control for electric vehicle (EV) charging in microgrids to solve voltage issues (Rastgoo et al., 2022) and the optimal placement of EV parking lots to improve energy management (Ahmadi et al., 2023). Further work explores comprehensive energy management systems that integrate photovoltaic and battery storage (Kermani et al., 2023) and the use of adaptive neural-fuzzy inference systems (ANFIS) to control EV charging based on grid frequency fluctuations (Mahdavi et al., 2024).

A recent study explored the benefits of integrating cloud infrastructure into robotic systems by Kehoe et al. (2015). Using big data access, cloud computing power, collective robot learning, and human computation, they demonstrate how the cloud facilitates more sophisticated and capable robotic applications. As a result of this survey, we can see how cloud-assisted technologies will play an increasingly important role in advancing robotics and automation in the future. By focusing on the evaluation and estimation of trust in cloud service providers (CSPs), Supriya, Sangeeta, and Patra have made significant contributions to the field of cloud computing trust management. In their work, they address the critical issue of trust in cloud computing environments, which is paramount to enterprises looking to adopt cloud services. To objectively compare the service levels of different CSPs, the authors introduced a model for Trust Management based on Fuzzy Logic in their 2013 study. To ensure that their applications comply with the required standards, this model assists customers in evaluating CSPs in terms of quality, reliability, and security. Fuzzy logic provides a nuanced approach to trust assessment, accommodating the inherent uncertainties and subjective perceptions of cloud computing trust. Direct experiences and reputational insights can significantly influence trust decisions in the dynamic and diverse marketplace of cloud services. During 2014, Supriya, Sangeeta, and Patra extended their investigation of trust management by focusing on the estimation of trust values within the context of Infrastructure as a Service (IaaS). In order to compare the trust values of various cloud service providers, they proposed a hierarchical trust model. The model measures the trustworthiness of CSPs using both numerical and non-numerical values, providing a more detailed framework for assessing cloud infrastructure reliability. A hierarchical approach allows CSPs to be evaluated more effectively, taking into account different aspects of trustworthiness that are critical for customers.

## 3. Results

Cloud Analyst is a simulation tool based on cloud computing modeling and analysis. Simulator parameters can be set in a wizard mode, and simulation results can be viewed in charts and tables. Using this simulator, we can repeat a scenario several times with different parameters and observe the results. In order to estimate traffic patterns and infrastructure needs, two methods were used: 1) Plain old mathematics with the theoretical maximum, and 2) Cloud Analyst traffic simulation software. A telecommunications company, Global NetSpan, has four branches in different parts of the world (Asia, Europe, Australia, and South America) and its application programs are hosted in data centers. The following three scenarios are considered:

1. Moving the entire infrastructure of the company to a single data center located in a specific geographic location, which has 100 virtual machines.

2. Designating two data centers in two geographic locations, each equipped with 50 virtual machines.

3. Establishing four data centers, each with 25 virtual machines.

The primary metric considered in this scenario is response time. The parameters considered in this simulation are simulated to assess their impact and interactions.

**Configuration**

**Host Configuration**

- Architecture: x86
- CPU: 2
- RAM: 2 GB
- Storage: 10 TB
- Bandwidth: 100000 Mbps

**Virtual Machine Configuration**

- VMM: XEN
- CPU: 1
- RAM: 512 MB
- Bandwidth: 1000 Mbps

**Cost**

- Virtual machine cost = 0.1 dollar
- Data transfer cost = 0.1 dollar
- Storage cost = 0.1 dollar

User grouping coefficient (number of users simulated in each UB) equals 1000. A request grouping coefficient of 100 represents the number of simulated requests. The executable instruction has a length of 250. We compare and evaluate three load balancing policies among virtual machines in a data center: Round Robin, Equally Spread, and Throttled. We have considered the nearest data center as the policy between the data center and the broker. 24 hours are set as the simulation duration. We can see a map of the world on the first page of the simulator. It has divided the world into six regions so that North America is in region 0, South America is in region 1, Asia is in region 3, Europe is in region 2, Africa is in region 4, and Australia is in region 5. The data center should be located in the same region as the branches, for example, if our company has four branches. In order to configure our scenario, we must click on the 'configure simulation' button. We can specify how long our simulation should run in the main tab; we choose 24 hours for our scenario. According to our scenario, our company has four branches. Each branch is a user base with a name and a location that can be anywhere in the world. Therefore, we need to add 4 user bases considering our 4 branches.

**Table 2:** Configuration setup for user bases and data center parameters in a cloud computing simulation spanning 20 hours

| Name | Region | Requests per User per Hr | Data Size per Request (bytes) | Peak Hours Start (GMT) | Peak Hours End (GMT) | Avg Peak Users | Avg Off-Peak Users |
|---|---|---|---|---|---|---|---|
| UB1 | 0 | 60 | 100 | 3 | 9 | 1000 | 100 |
| UB2 | 3 | 60 | 100 | 3 | 9 | 1000 | 100 |
| UB3 | 2 | 60 | 100 | 3 | 9 | 1000 | 100 |
| UB4 | 5 | 60 | 100 | 3 | 9 | 1000 | 100 |

Table 2 shows a configuration setup for a 20-hour simulation detailing the parameters for four distinct user bases across various global regions. First user base (UB1) is in South America (Region 1), second user base (UB2) is in Asia (Region 3), third user base (UB3) is in Europe (Region 2), and fourth user base (UB4) is in Australia (Region 5). A consistent pattern of 60 requests per user per hour is configured for each user base, with a data size of 100 bytes per request. There are an average of 1000 users during peak hours and 100 users off-peak, with peak hours defined as 3 GMT to 9 GMT. Moreover, on this page, one can configure the deployment settings for the application, including the data center specifications. In the beginning, a single entity hosts 100 virtual machines in the data center. Virtual machines have 512 MB of RAM and 1000 Mbps of bandwidth.

**Table 3:** Application Deployment Configuration

| Service Broker Policy | Data Center | # VMs | Image Size | Memory | Bandwidth |
|---|---|---|---|---|---|
| Closest Data Center | DC1 | 100 | 10000 | 512 | 1000 |
| Optimise Response Time | DC1 | 100 | 10000 | 512 | 1000 |

The application deployment configuration section of a cloud computing simulation interface is shown in Table 3. Configuration interface showcasing the application deployment settings for a data center in a cloud computing simulation, with the ability to choose between 'Closest Data Center' and 'Optimize Response Time' as service broker policies in table 3. This section allows users to define the characteristics of their data centers. A service broker policy can be configured based on two different strategies: 'Closest Data Center' and 'Optimize Response Time'. Both strategies will be simulated and compared. Additionally, users can specify the configuration of their data center(s) within the data center configuration tab. The number of data centers and the hardware in each data center are included in this. In the first step of the simulation, there is a single data center with 100 virtual machines (VMs), an image size of 10,000, 512 MB of memory, and 1000 Mbps of bandwidth. There are two CPUs in the data center in the scenario. The data center configuration consists of DC1, located in region 0, operating on x86 architecture with Linux as the operating system and Xen as the virtual machine monitor, with a VM cost of $0.10 per hour, memory cost of $0.05 per second, storage cost of $0.10 per second, data transfer cost of $0.10 per GB, and a total of two physical hardware units. Table 2 shows configuration of data center DC1, including cost parameters and physical hardware specifications, used for evaluating load balancing policies in the cloud simulation environment.

**Table 4**: Physical Hardware Details of Data Center DC1

| HW Unit ID | Memory (MB) | Storage (MB) | Available BW | Processors | Processor Speed | VM Policy |
|---|---|---|---|---|---|---|
| 0 | 204,800 | 100,000,000 | 1,000,000 | 4 | 10,000 | TIME_SHARED |
| 1 | 204,800 | 100,000,000 | 1,000,000 | 4 | 10,000 | TIME_SHARED |

A cloud simulation platform is used to simulate a data center, DC1, which is shown in Table 4. Located in region 0, DC1 utilizes an x86 architecture, Linux operating system, and Xen virtual machine monitor (VMM). According to the cost metrics associated with this data center, virtual machines are charged $0.1 per hour, memory is charged $0.05 per second, storage is charged $0.1, and data transfer is charged $0.1. In addition, DC1 contains two physical hardware units. Two units with considerable resources are revealed by the physical hardware details for DC1. The units have 204,800 MB of memory and 10,000,000 MB of storage, with a bandwidth of 100,000,000 Mbps available. These units are equipped with four processors each, which operate at a speed of 10,000 MIPS (Million Instructions Per Second), and adhere to a 'TIME_SHARED' virtual machine (VM) policy that allows multiple VMs to share the same physical hardware resources in a time-efficient manner. To simulate users, users must define the user grouping coefficient in the advanced tab of the simulation platform, which is set to 1000. The request grouping coefficient is set to 100 to reflect the number of requests per group, and the job queue length is 250. Moreover, the load balancing policy is determined, with strategies 1 to 3 (Round Robin, Equally Spread, and Throttled) uniformly applied and compared across the three approaches.

**Table 5:** User and Request Configuration Under Different Load Balancing Policies

| **Parameter** | **Round Robin** | **Equally Spread Current Execution Load** | **Throttled** |
|---|---|---|---|
| User grouping factor (users per user base) | 1000 | 1000 | 1000 |
| Request grouping factor (requests per application server) | 100 | 100 | 100 |
| Executable instruction length per request (bytes) | 250 | 250 | 250 |
| Load balancing policy | Round Robin | Equally Spread Current Execution Load | Throttled |

Table 5 shows a segment of a cloud computing simulation interface where various parameters are set for load balancing. In these configurations, the user grouping factor in User Bases, which represents the number of simultaneous users from one user base, is uniformly set to 1000. Data Centers consistently set the request grouping factor to 100, which represents the number of simultaneous requests a single application server instance can handle. Additionally, 250 bytes are specified as the executable instruction length per request. Across virtual machines within a single data center, each subfigure illustrates a different load balancing policy. The first policy illustrated is Round Robin, which distributes tasks evenly, in a cyclical order, without prioritization, to the next available node. Equally Spread Current Execution Load aims to distribute the load equally among all nodes, ensuring that no single node is overloaded. Lastly, Throttled allocates tasks dynamically to manage the load efficiently across servers, often reducing the number of active connections to a given server to prevent overloading. Furthermore, each region's delay and bandwidth parameters can be specified through the defined internet option within the simulation. For simulating real-world internet conditions, these parameters are crucial, affecting cloud service performance and load balancing effectiveness.

**Table 6:** Transmission Delay Matrix Between Regions (ms)

| Region | 0 | 1 | 2 | 3 | 4 | 5 |
|---|---|---|---|---|---|---|
| **0** | 25 | 100 | 150 | 250 | 250 | 100 |
| **1** | 100 | 25 | 250 | 500 | 350 | 200 |
| **2** | 150 | 250 | 25 | 150 | 150 | 200 |
| **3** | 250 | 500 | 150 | 25 | 500 | 500 |
| **4** | 250 | 350 | 150 | 500 | 25 | 500 |
| **5** | 100 | 200 | 200 | 500 | 500 | 25 |

**Table 7:** Available Bandwidth Matrix Between Regions (Mbps)

| Region | 0 | 1 | 2 | 3 | 4 | 5 |
|---|---|---|---|---|---|---|
| **0** | 2000 | 1000 | 1000 | 1000 | 1000 | 1000 |
| **1** | 1000 | 800 | 1000 | 1000 | 1000 | 1000 |
| **2** | 1000 | 1000 | 2500 | 1000 | 1000 | 1000 |
| **3** | 1000 | 1000 | 1000 | 1500 | 1000 | 1000 |
| **4** | 1000 | 1000 | 1000 | 1000 | 500 | 1000 |
| **5** | 1000 | 1000 | 1000 | 1000 | 1000 | 2000 |

In Tables 6 and 7, two matrices are shown within a cloud computing simulation. Table 6 presents the inter-regional network latency values, in milliseconds, used to model transmission delays between different regions in the cloud simulation environment. Table 7 shows the available inter-regional bandwidth capacities, in Mbps, applied in the simulation to represent data transfer capabilities between regions. First, there is the Delay Matrix, which outlines the transmission delay between various regions in milliseconds. The second matrix shows the bandwidth available between regions for the simulated application, in megabits per second (Mbps). The Delay Matrix shows the latency involved when transmitting data between regions. The delay between Region 0 and Region 1 is 100 milliseconds, whereas the delay between Region 0 and Region 3 is 250 milliseconds. For time-sensitive applications, this matrix is vital for understanding the communication delays in the simulated cloud network. In the Bandwidth Matrix, data transfer capacity between regions is shown. High bandwidth values suggest a higher capacity for data flow, which can improve performance and reduce data transfer times. Region 2 and Region 3 have a bandwidth of 2,500 Mbps, indicating a higher capacity for data transmission than Region 4 and Region 5, which have a bandwidth of 500 Mbps. Simulating cloud computing network characteristics requires these matrices. To simulate real-world conditions and understand how different network configurations can affect cloud service performance, they provide the necessary parameters.

The comment at the bottom indicates the intent to simulate the initial state of work and proceed by clicking on "run simulation," which indicates that the user is about to initiate the simulation process using the defined network parameters to analyze and evaluate the performance of a cloud-based application or service.

### *3.1. Result of strategy step 1*

To evaluate the impact of different load balancing strategies on system performance, a cloud simulation was conducted using three widely adopted policies: Round Robin, Equally Spread Current Execution Load, and Throttled. Key performance indicators, including overall response time and data center processing time, were recorded to assess both user-perceived latency and internal system efficiency. These metrics provide insight into how effectively each policy distributes workload across virtual machines under identical configuration settings. Table 8 summarizes the average, minimum, and maximum response times, along with corresponding data center processing times, for each load balancing policy. The results show that the Round Robin policy achieved the lowest average overall response time, while the Equally Spread Current Execution Load and Throttled policies demonstrated reduced data center processing times, indicating more balanced internal resource utilization. However, the Throttled policy resulted in the highest average response time, reflecting stricter VM allocation constraints.

**Table 8**: Overall Response Time Comparison Under Different Load Balancing Policies

| Load Balancing Policy | Avg. Response Time (ms) | Min. Response Time (ms) | Max. Response Time (ms) | Avg. Data Center Processing Time (ms) | Min. Processing Time (ms) | Max. Processing Time (ms) |
|---|---|---|---|---|---|---|
| Round Robin | 264.48 | 39.81 | 664.84 | 2.18 | 0.03 | 8.31 |
| Equally Spread Current Execution Load | 265.69 | 39.81 | 664.84 | 1.61 | 0.03 | 3.23 |
| Throttled | 269.09 | 44.99 | 664.84 | 1.62 | 0.03 | 3.18 |

In Table 8, we compare the performance of three different load balancing strategies: Round Robin, Equally Spread, and Throttled. Using the Round Robin strategy (A), the average overall response time was 264.48 milliseconds (ms), while the minimum and maximum response times ranged from 39.81 ms to 664.84 ms. As a result, the data center processing time averaged 2.18 ms, indicating efficient handling within the data center. The Equally Spread strategy (B) resulted in a slight increase in the average overall response time to 265.69 ms, with the same minimum and maximum response times as Round Robin. However, the data center processing time improved significantly, dropping to an average of 1.61 milliseconds. The throttled strategy (B) resulted in an average overall response time of 269.09 milliseconds. Minimum response time increased to 44.99 ms, while maximum response time remained the same. It took 1.62 milliseconds on average to process data in the data center, comparable to the Equally Spread strategy. In spite of the marginally higher overall response time of the Throttled strategy, the data center processing times are comparable to those of the Equally Spread strategy, suggesting Throttled may manage task distribution with a bit more caution, possibly in order to prevent resource overload. As a result, Round Robin and Equally Spread offer similar overall performance, with Equally Spread having a slight edge in terms of data center processing efficiency.

**Table 9**: Result of Round Robin with strategy step 1

| Userbase | Avg (ms) | Min (ms) | Max (ms) |
|---|---|---|---|
| UB1 | 52.12 | 39.81 | 67.68 |
| UB2 | 502.22 | 377.19 | 664.84 |
| UB3 | 301.31 | 233.19 | 359.70 |
| UB4 | 201.85 | 153.45 | 255.42 |

**Table 10:** Result of Equally Spread with strategy step 1

| Userbase | Avg (ms) | Min (ms) | Max (ms) |
|---|---|---|---|
| UB1 | 51.61 | 39.81 | 60.43 |
| UB2 | 507.80 | 424.87 | 664.84 |
| UB3 | 297.20 | 255.12 | 357.30 |
| UB4 | 201.01 | 169.29 | 228.42 |

**Table 11:** Result of Throttled with strategy step 1

| Userbase | Avg (ms) | Min (ms) | Max (ms) |
|---|---|---|---|
| UB1 | 51.97 | 44.99 | 58.31 |
| UB2 | 515.07 | 442.82 | 664.84 |
| UB3 | 291.49 | 266.42 | 339.57 |
| UB4 | 201.56 | 184.14 | 215.14 |

Based on strategy step 1 in the cloud computing simulation, Tables 9, 10, and 11 summarize the response time results across four user bases (UB1, UB2, UB3, and UB4) under three different load balancing strategies: Round Robin, Equally Spread, and Throttled, respectively. As shown in Table 9, the Round Robin strategy yields the lowest average response time for UB1 at 52.12 ms, while UB3 experiences the highest average response time at 301.31 ms. For UB1 and UB2, response times range from 39.81 ms to 233.19 ms, whereas for UB3 and UB4, response times extend up to 664.84 ms, indicating higher latency under heavier workloads. The results for the Equally Spread strategy are presented in Table 10. Similar to Round Robin, UB1 exhibits the shortest average response time at 51.61 ms, while UB3 records the highest average response time at 297.20 ms. This strategy maintains the same minimum response times as Round Robin but achieves slightly lower maximum response times, with UB3's maximum response time decreasing from 359.70 ms (Table 9) to 357.30 ms (Table 10).

Table 11 reports the performance of the Throttled strategy. While UB1 and UB4 show response times comparable to the other strategies, UB2 and UB3 experience increased average response times of 515.07 ms and 291.49 ms, respectively. However, under the Throttled strategy, UB3's maximum response time is reduced to 339.57 ms, representing an improvement compared to the previous strategies. he results across Tables 9–11 indicate that Round Robin and Equally Spread strategies deliver similar performance with minor variations, whereas the Throttled strategy produces mixed outcomes reducing maximum response times for some user bases while increasing average response times for others. Therefore, the choice of load balancing strategy should be guided by specific performance objectives and the workload characteristics of the cloud computing environment.

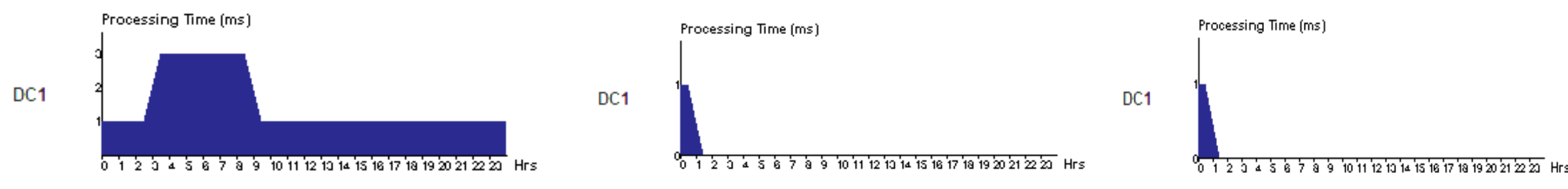

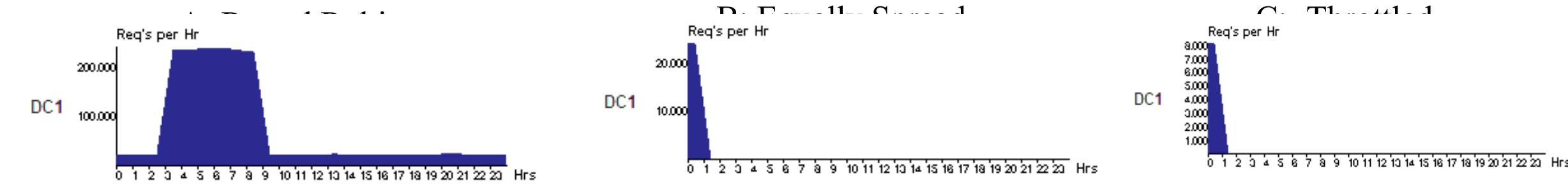

A: Round Robin B: Equally Spread C: Throttled

**Fig. 1** Histograms depicting processing times at DC1 under different load balancing strategies, illustrating the distribution of computational load over a 24-hour period.

**Fig. 2** Histograms showing the number of requests processed per hour at DC1 under different load balancing strategies, reflecting the request handling capacity and operational load

Figure 1 shows three histograms representing the processing times for DC1 under Round Robin (A), Equally Spread (B), and Throttled (C). The histograms show how long it takes the data center to process requests over a 24-hour period. Figure 2shows the number of requests handled by DC1 per hour under the same strategies. It is important to assess the performance and efficacy of each load balancing strategy implemented by examining the loading histograms for each data center throughout the simulation period. These histograms represent the distribution and intensity of computational demand placed on the data centers.

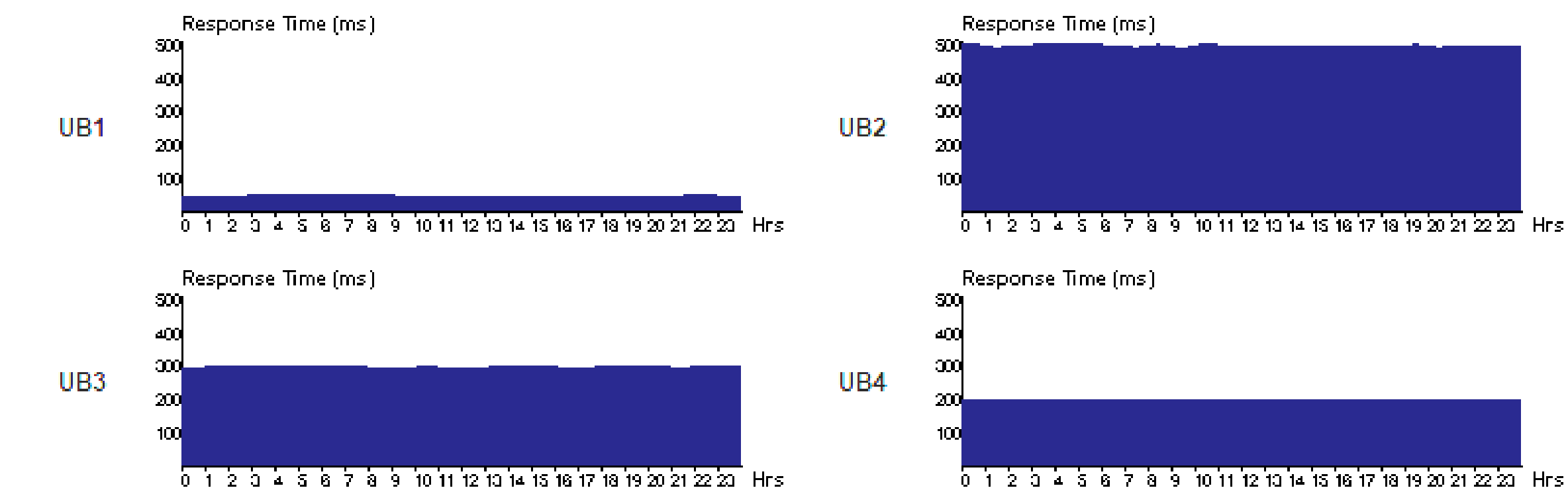

A: Round Robin

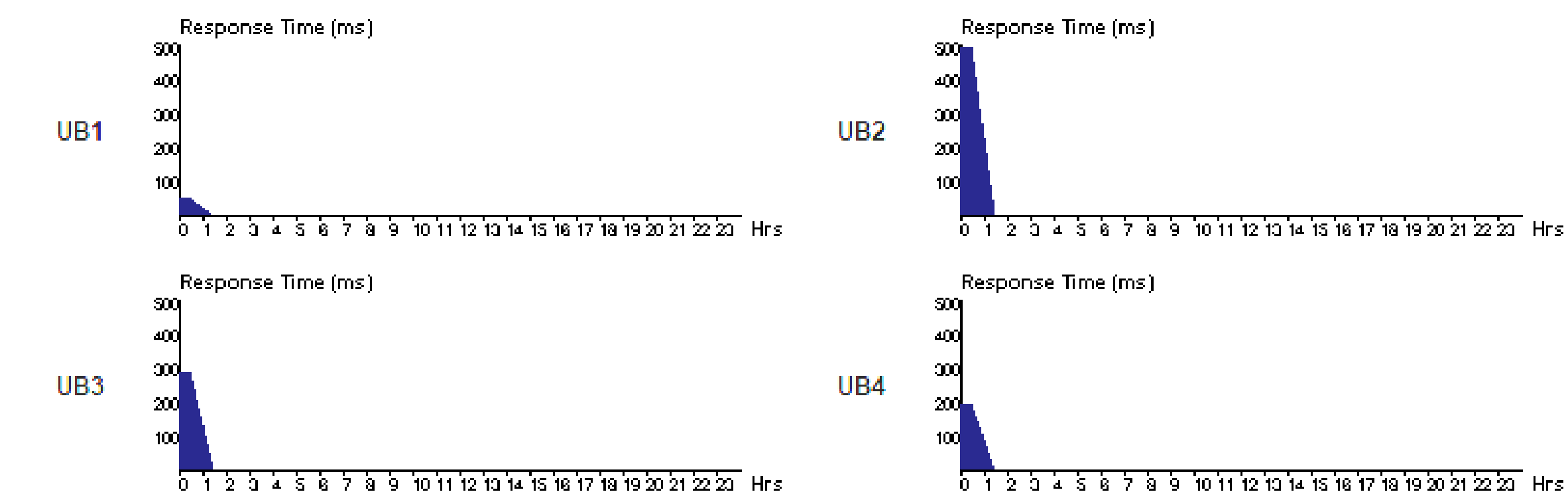

B: Equally Spread

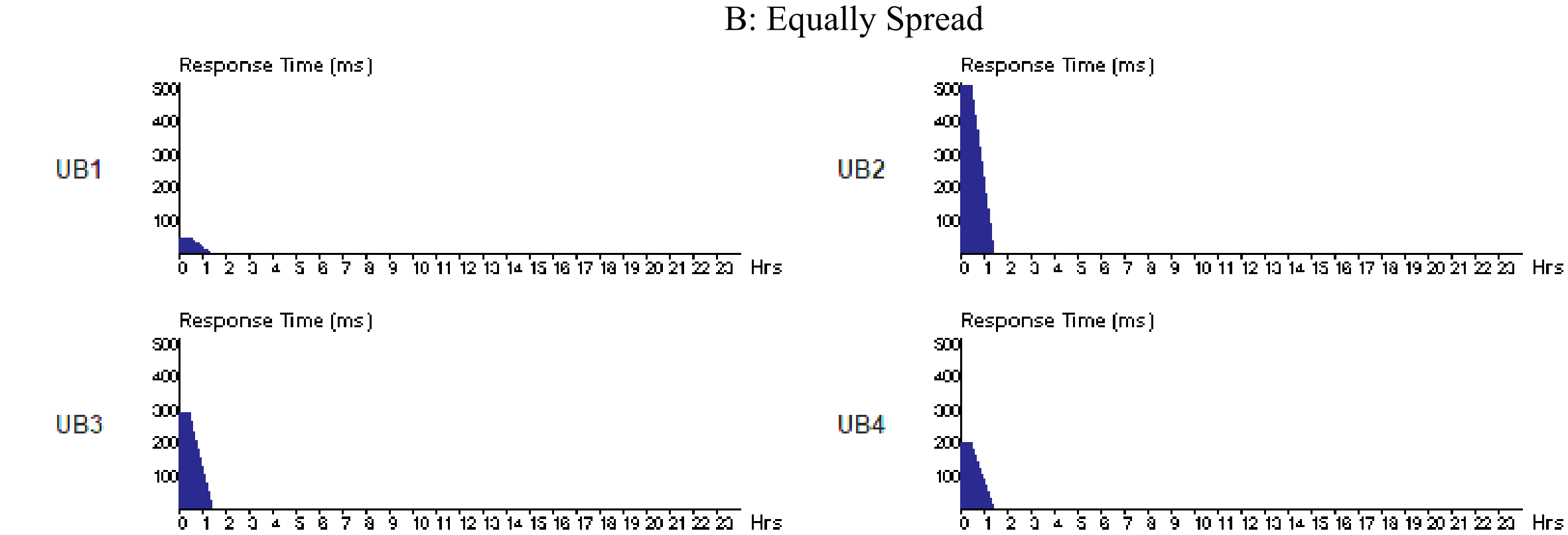

C: Throttled

**Fig. 3** A comparative display of user base hourly response times under three load balancing strategies, highlighting the performance of the "Closest Data Center" service broker policy during strategy step 1 in a cloud computing environment.

Figure 3 shows the hourly response times for four different user bases (UB1, UB2, UB3, UB4) under three different load balancing strategies: Round Robin, Equally Spread, and Throttled. Based on the "Closest Data Center" as the service broker policy, these graphs are part of a simulation analysis for strategy step 1. In the Round Robin strategy (A), the histograms show the distribution of response times over a 24-hour period. A relatively uniform distribution of load is achieved by distributing incoming requests sequentially and evenly across all available servers. With Equally Spread (B), incoming requests are distributed evenly among the servers, ensuring that no one server bears too much load at any given time. Response time histograms for this strategy would typically show a more balanced distribution of response times across the hours, reflecting the strategy's goal of maintaining equilibrium. To prevent overloading, the Throttled strategy (C) limits the number of active connections or requests to a server at any given time. When the throttling mechanism kicks in to prevent server overload, the response time distribution would show spikes and smoother periods when the load is within acceptable limits. Load balancing strategies are visualized and compared using these histograms. Stakeholders can analyze response time distributions to determine which strategy will best meet their operational needs, taking into account factors like user demand patterns, server capacity, and the overall goal of maintaining a high quality of service.

**Table 12:** Result of Cost for Round Robin with strategy step 1 (Closet Data Center)

| Data Center | VM Cost | Data Transfer Cost | Total |
|---|---|---|---|
| DC1 | 12.001 | 17.803 | 29.804 |

**Table 13:** Result of Cost for Equally Spread with strategy step 1 (Closet Data Center)

| Data Center | VM Cost | Data Transfer Cost | Total |
|---|---|---|---|
| DC1 | 0.502 | 0.233 | 0.735 |

**Table 14:** Result of Cost for Throttled with strategy step 1 (Closet Data Center)

| Data Center | VM Cost | Data Transfer Cost | Total |
|---|---|---|---|
| DC1 | 0.168 | 0.078 | 0.246 |

Based on the "Closest Data Center" policy for strategy step 1, tables 12,13 and 14 summarize the estimated costs associated with running a cloud computing simulation using three different load balancing strategies: Round Robin, Equally Spread, and Throttled. The following table shows the costs associated with the use of virtual machines (VMs) and data transfers within DC1. We note that the VM cost is 12.001, the data transfer cost is 17.803, and the total cost is 29.804. Under the Equally Spread strategy, the same data center has a significantly lower cost profile. The VM cost is reduced to 0.502, the data transfer cost to 0.233, and the total cost is 0.735. The Throttled strategy has the lowest costs of the three strategies. VM cost is 0.168, data transfer cost is 0.078, and total cost is 0.246. To evaluate the economic efficiency of load balancing strategies, these cost estimations are crucial. Clearly, the Throttled strategy is the most cost-effective, and the Equally Spread strategy is significantly cheaper than the Round Robin strategy. With these figures, cloud service providers and users can make informed decisions about which load balancing strategy to employ, depending on both their performance needs and budgets. "Cost" in the simulation provides users with an idea of the financial implications of their simulated cloud infrastructure and service usage, such as the costs associated with virtual machine operation and data transfer.

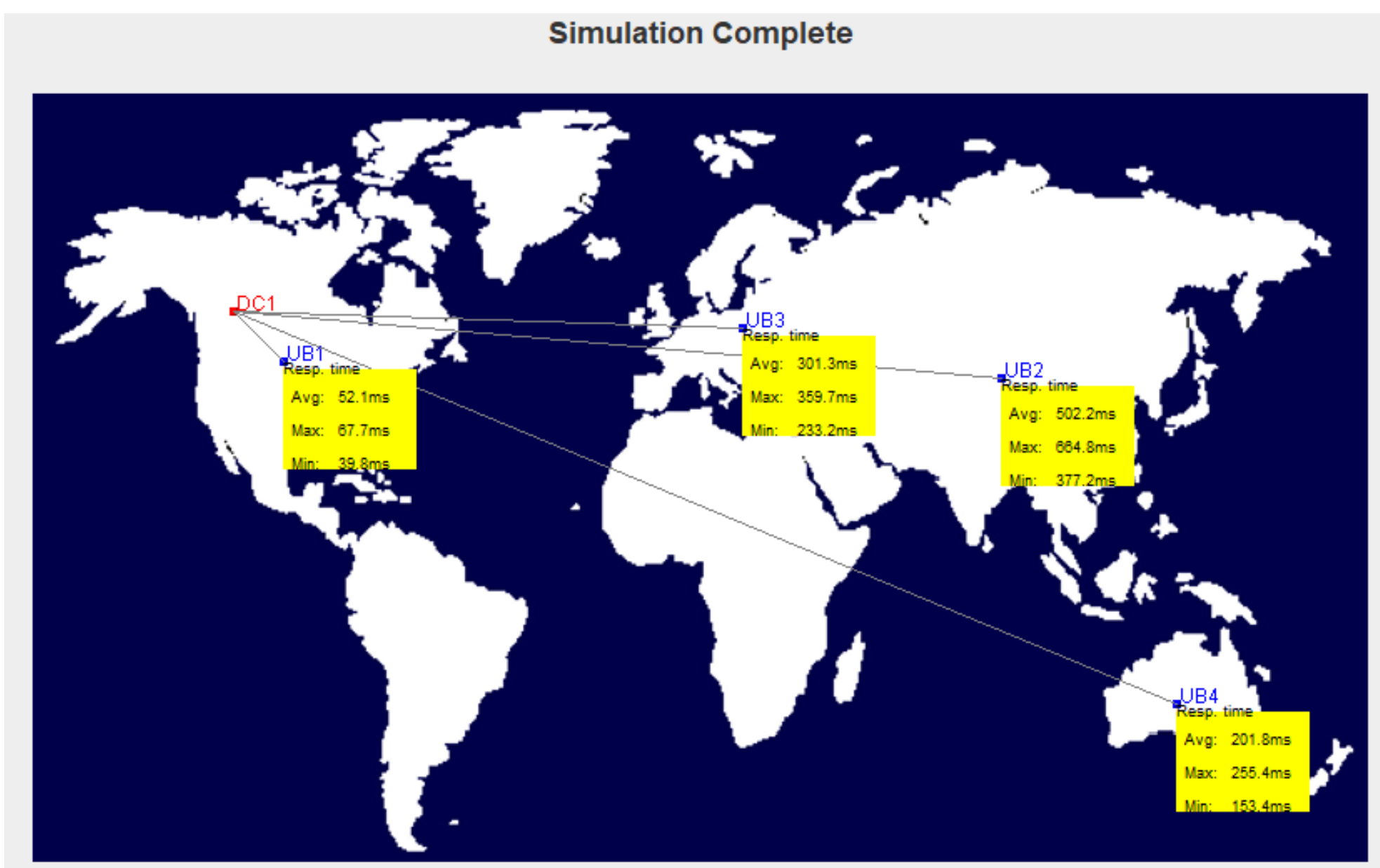

**Fig. 4** Graphical representation of a cloud computing simulation showing user base response times relative to a central data center, illustrating the geographical aspects of network latency and performance.(Load Balancing technique Round Robin with Strategy step 1)

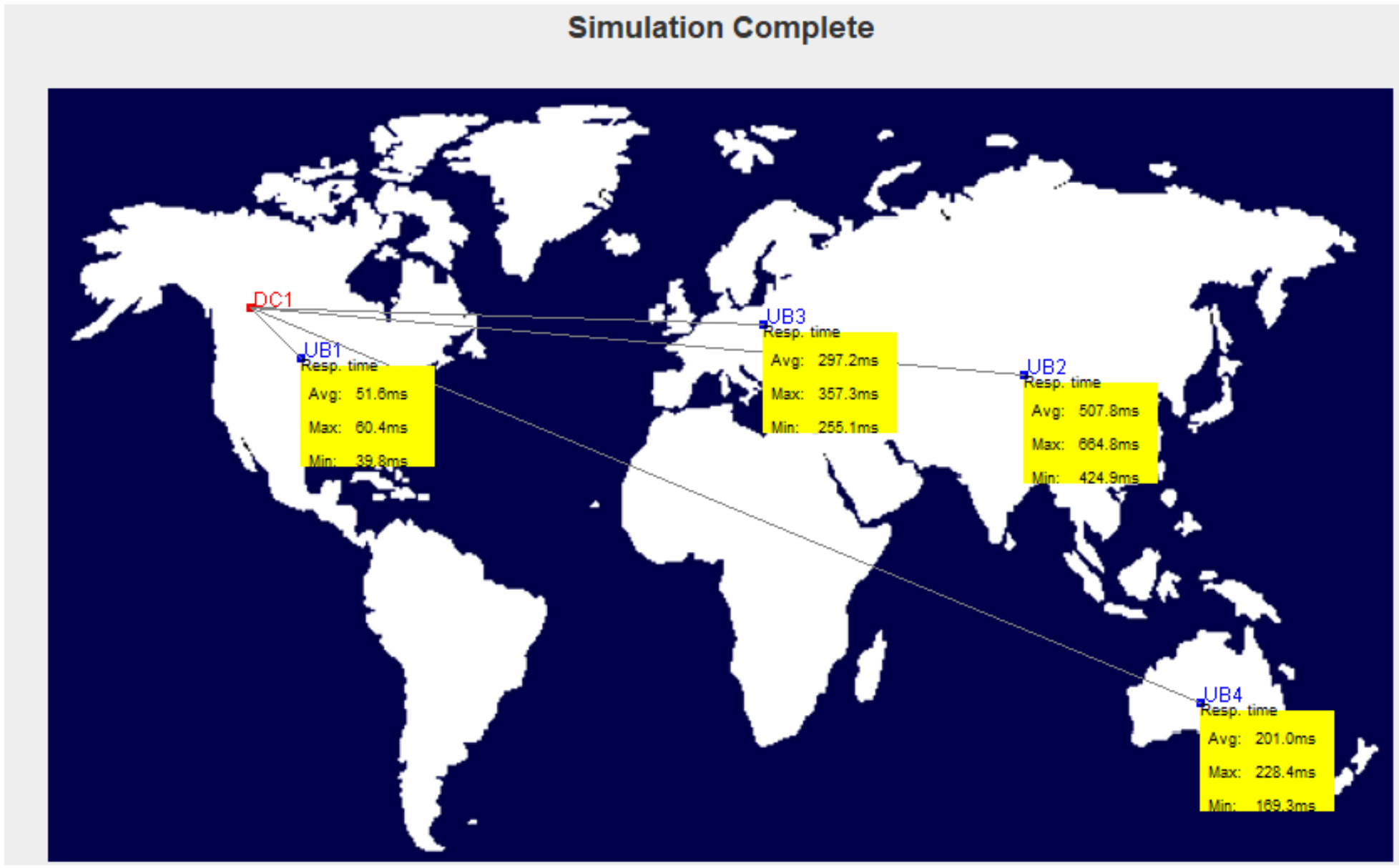

**Fig. 5** Graphical representation of a cloud computing simulation showing user base response times relative to a central data center, illustrating the geographical aspects of network latency and performance.( Load Balancing technique Equally Spread with Strategy step 1)

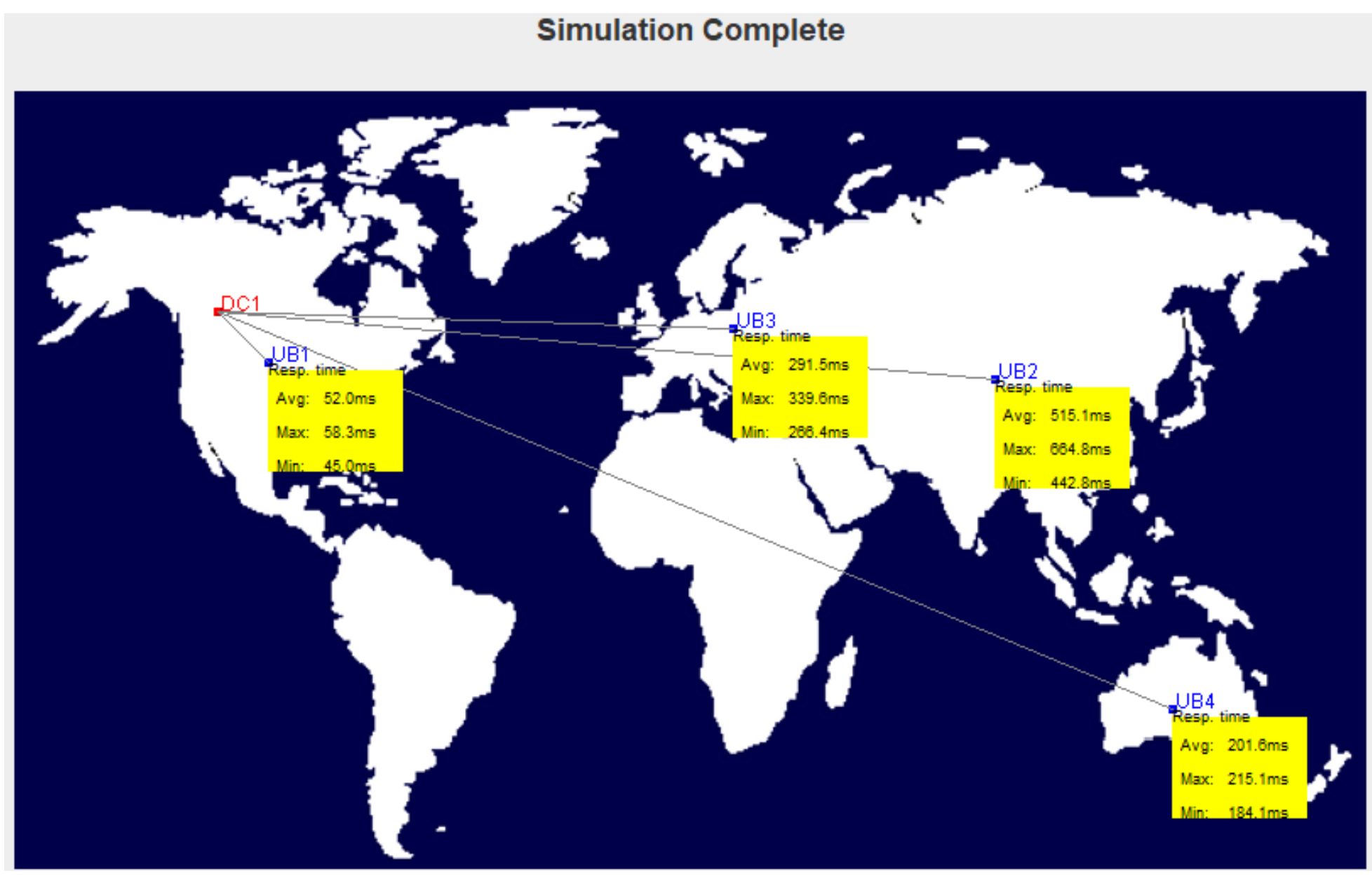

**Fig. 6** Graphical representation of a cloud computing simulation showing user base response times relative to a central data center, illustrating the geographical aspects of network latency and performance.( Load Balancing technique Throttled with Strategy step 1)

An output of a cloud computing simulation is shown in figures 10, 11 and 12. Data centers and user bases (UB1, UB2, UB3, UB4) are indicated on a world map with corresponding response time metrics. One can view the average response times across the entire simulation, as well as the maximum and minimum response times. Additionally, the average, maximum, and minimum response times for each region are displayed, providing details regarding the performance metrics for each user base in relation to the data center (DC1). A quick assessment of the latency between the user bases and the data center can be made with this visual representation. As an example, UB1 has the lowest average response time, suggesting a closer proximity or better connectivity to DC1, while UB2 has the highest average response time, suggesting a greater distance or network delay. Furthermore, UB3 and UB4 show their respective response times, which can be used to evaluate the efficiency of data transfer on the cloud network.

### *3.2. Result of strategy step 2*

To further evaluate the performance of different load balancing strategies under an alternative configuration, strategy step 2 was implemented in the cloud simulation environment. The Round Robin, Equally Spread, and Throttled policies were assessed using key performance metrics, including overall response time and data center processing time. These metrics provide insight into both end-user experience and internal processing efficiency under varying workload distribution mechanisms.

Table 13 presents a comparative summary of the average, minimum, and maximum overall response times, as well as data center processing times, for each load balancing strategy. The results indicate that the Round Robin policy achieves the lowest average overall response time (264.48 ms) and substantially lower processing overhead compared to the other strategies. In contrast, both the Equally Spread and Throttled policies exhibit higher average response times (271.87 ms and 271.88 ms, respectively) and significantly increased data center processing times, reflecting greater computational overhead under these policies. Although the minimum response times remain similar across strategies, the maximum response time increases slightly for the Equally Spread and Throttled approaches. Overall, the findings suggest that while Equally Spread and Throttled strategies may offer improved workload control, they do so at the cost of increased processing delay, highlighting a trade-off between resource management and response time performance.

**Table 13:** Comparative summary of overall response times across three load balancing strategies at strategy step 2, highlighting differences in average, minimum, and maximum response times as well as data center processing times

| Load Balancing Policy | Avg. Response Time (ms) | Min. Response Time (ms) | Max. Response Time (ms) | Avg. Data Center Processing Time (ms) | Min. Processing Time (ms) | Max. Processing Time (ms) |
|---|---|---|---|---|---|---|
| Round Robin | 264.48 | 39.81 | 664.84 | 2.18 | 0.03 | 8.31 |
| Equally Spread | 271.87 | 41.39 | 677.05 | 9.65 | 0.16 | 16.44 |
| Throttled | 271.88 | 41.39 | 677.05 | 9.65 | 0.16 | 16.44 |

In Table 13, the overall response times are summarized from a cloud computing simulation comparing three different load balancing strategies at strategy step 2. The Round Robin strategy (A) shows an average overall response time of 264.48 milliseconds (ms), with minimum and maximum response times of 39.81 and 664.84 ms, respectively. The data center processing time is notably efficient, averaging at 2.18 ms. In the Equally Spread strategy (B), there is a slight increase in the average overall response time to 271.87 ms. The minimum and maximum response times show little change from the Round Robin strategy, and the data center processing time shows a considerable increase to an average of 9.65 ms. The Throttled strategy (C) also results in an average overall response time of 271.88 ms, mirroring the response time of the Equally Spread strategy. Averaging 9.65 milliseconds in data center processing time, the Equally Spread strategy remains consistent. Despite the fact that the Round Robin strategy has a slightly lower overall response time than the Equally Spread and Throttled strategies, it has a significantly faster data center processing time than the Equally Spread and Throttled strategies. Round Robin may be more efficient within the data center, but Equally Spread and Throttled may be competitive when considering network latency and other factors.

**Table 14:** Result of Round Robin with strategy step 2

| Userbase | Avg (ms) | Min (ms) | Max (ms) |
|---|---|---|---|
| UB1 | 60.64 | 39.81 | 95.57 |
| UB2 | 510.91 | 395.64 | 691.58 |
| UB3 | 309.59 | 233.19 | 386.08 |
| UB4 | 210.02 | 158.33 | 267.39 |

**Table 15**: Result of Equally Spread with strategy step 2

| Userbase | Avg (ms) | Min (ms) | Max (ms) |
|---|---|---|---|
| UB1 | 59.46 | 41.39 | 80.42 |
| UB2 | 510.05 | 395.33 | 677.05 |
| UB3 | 308.56 | 237.25 | 371.55 |
| UB4 | 209.20 | 155.68 | 264.82 |

**Table 16:** Result of Throttled with strategy step 2

| Userbase | Avg (ms) | Min (ms) | Max (ms) |
|---|---|---|---|
| UB1 | 59.47 | 41.39 | 80.42 |
| UB2 | 509.94 | 395.33 | 677.05 |
| UB3 | 308.53 | 237.25 | 371.55 |
| UB4 | 209.24 | 155.68 | 264.82 |

Tables 14–16 present the response time performance of the Round Robin, Equally Spread, and Throttled load balancing strategies under strategy step 2 across four user bases (UB1–UB4). Overall, all three strategies exhibit similar performance patterns, with UB1 consistently achieving the lowest average response times (approximately 59–61 ms) and UB2 experiencing the highest average response times (approximately 509–511 ms) due to higher workload intensity. The Equally Spread and Throttled strategies demonstrate slightly improved average and maximum response times for UB1, UB3, and UB4 compared to Round Robin, indicating more balanced workload distribution across virtual machines. In contrast, response times for UB2 remain high across all strategies, suggesting that heavy user demand dominates scheduling effects. Minimum

response times remain comparable among the three approaches, while the Equally Spread and Throttled strategies achieve modest reductions in maximum response times for UB3 and UB4, reflecting improved peak-load handling under strategy step 2. UB1 experiences the quickest average response time of 60.64 milliseconds (ms), suggesting that the Round Robin strategy may be effective for users in this base. Conversely, UB2 has the longest average response time of 510.91 ms, possibly due to a heavier load or a less optimal distribution of requests for this user base. UB1 again showed the shortest average response time at 59.46 ms for Equally Spread with strategy step 2, marginally better than Round Robin. The response time of UB2 remains high, but has improved slightly compared to the Round Robin strategy at 510.05 ms, suggesting that the Equally Spread strategy may offer a slight performance boost. Throttled with strategy step 2 shows similar response times to Equally Spread. While UB1 maintains a low average response time of 59.47 milliseconds, UB2's average remains over 500 milliseconds, closely matching the Equally Spread strategy. Under the conditions set for strategy step 2, Equally Spread and Throttled strategies may perform similarly.

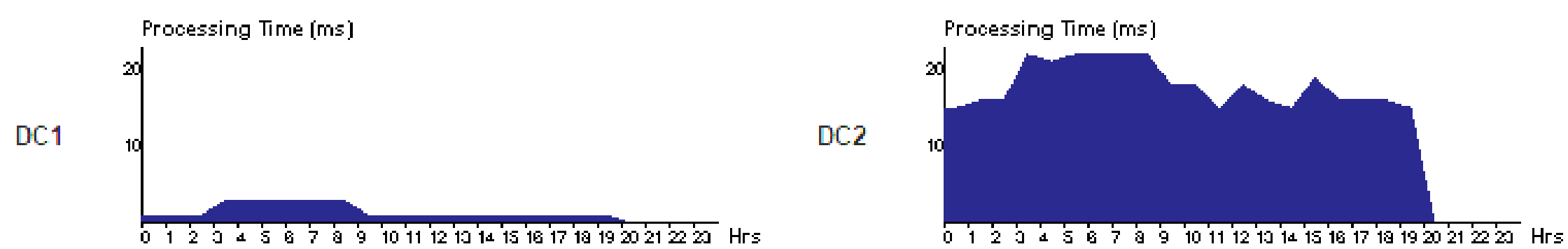

A: Round Robin

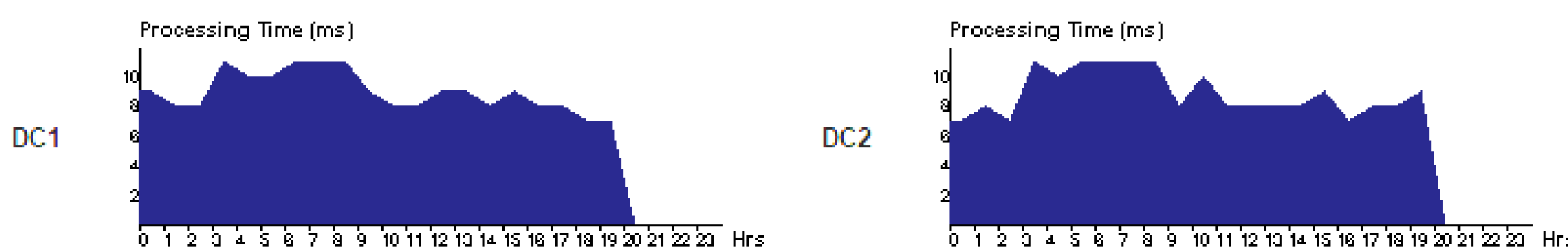

B: Equally Spread

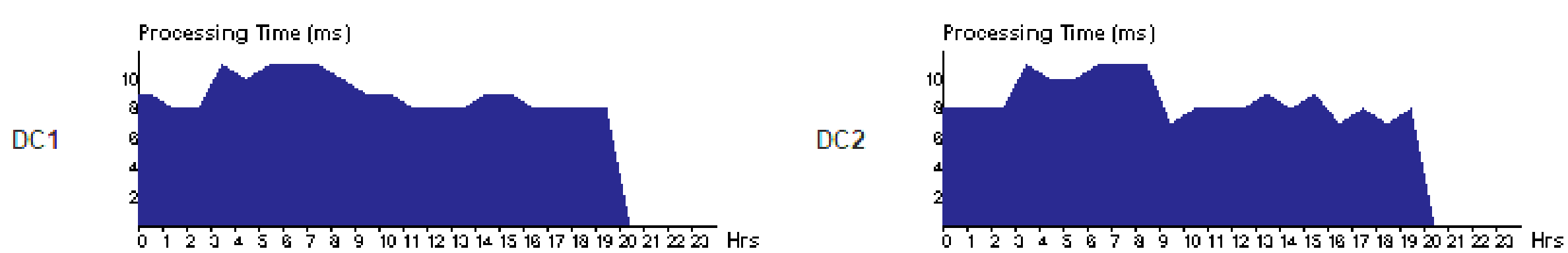

C: Throttled

**Fig. 7** processing time for strategy step 2

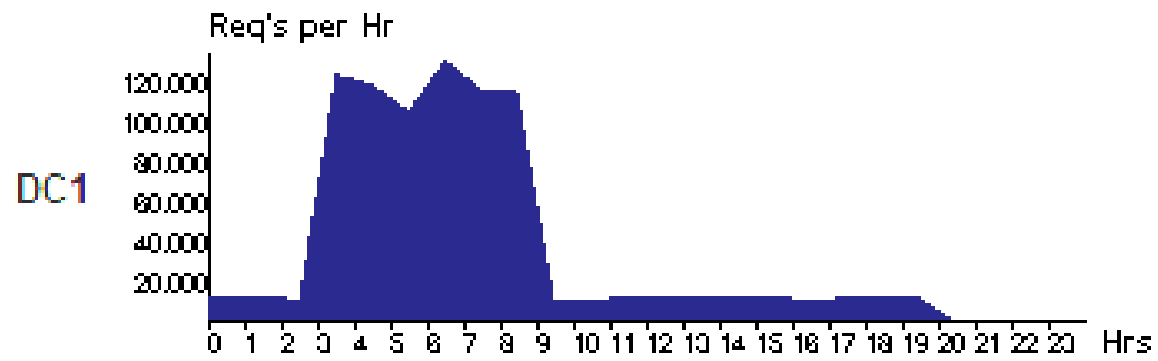

A: Round Robin

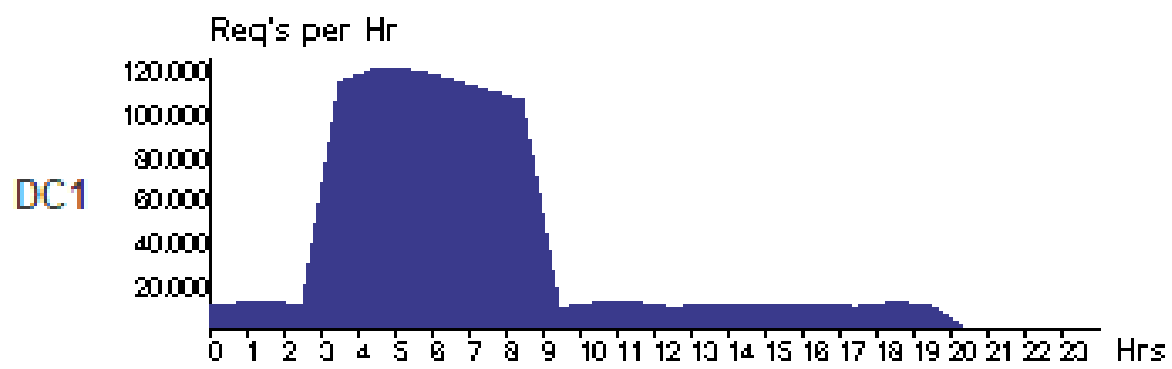

B: Equally Spread

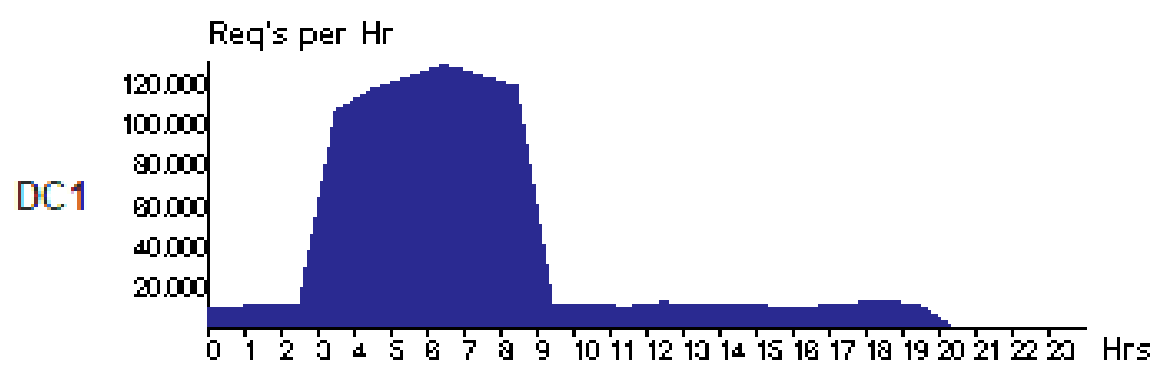

C: Throttled

**Fig. 8** Request per hr for strategy step 2

Round Robin strategy (A) shows fluctuations in processing time throughout the day, with peaks associated with periods of high traffic. Round Robin distributes requests cyclically, which can result in uneven processing times when demand surges. With the Equally Spread strategy (B), processing times for both data centers appear to be more stable, though they are still subject to variability. In comparison to Round Robin, this strategy distributes the load evenly across all servers, resulting in more uniform processing times. For DC1 and DC2, the Throttled strategy (C) suggests a controlled handling of requests in order to prevent server overload. It appears that throttling may effectively manage the load to maintain consistent processing times, since the times are generally lower and exhibit less variance.

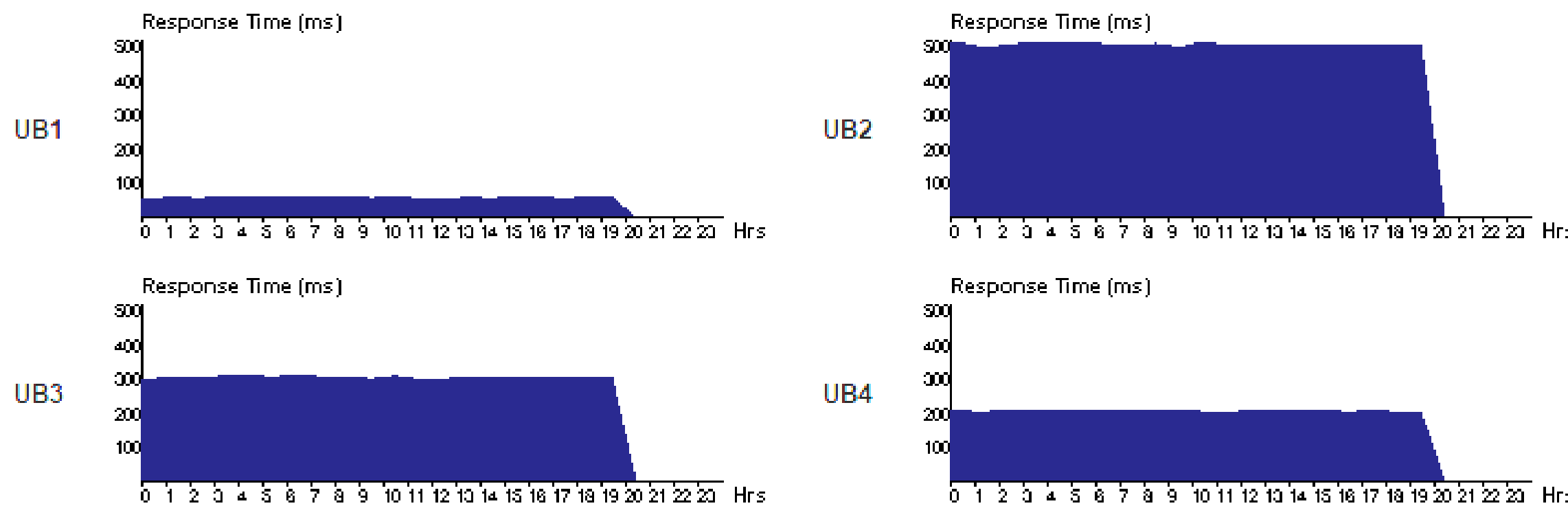

A: Round Robin

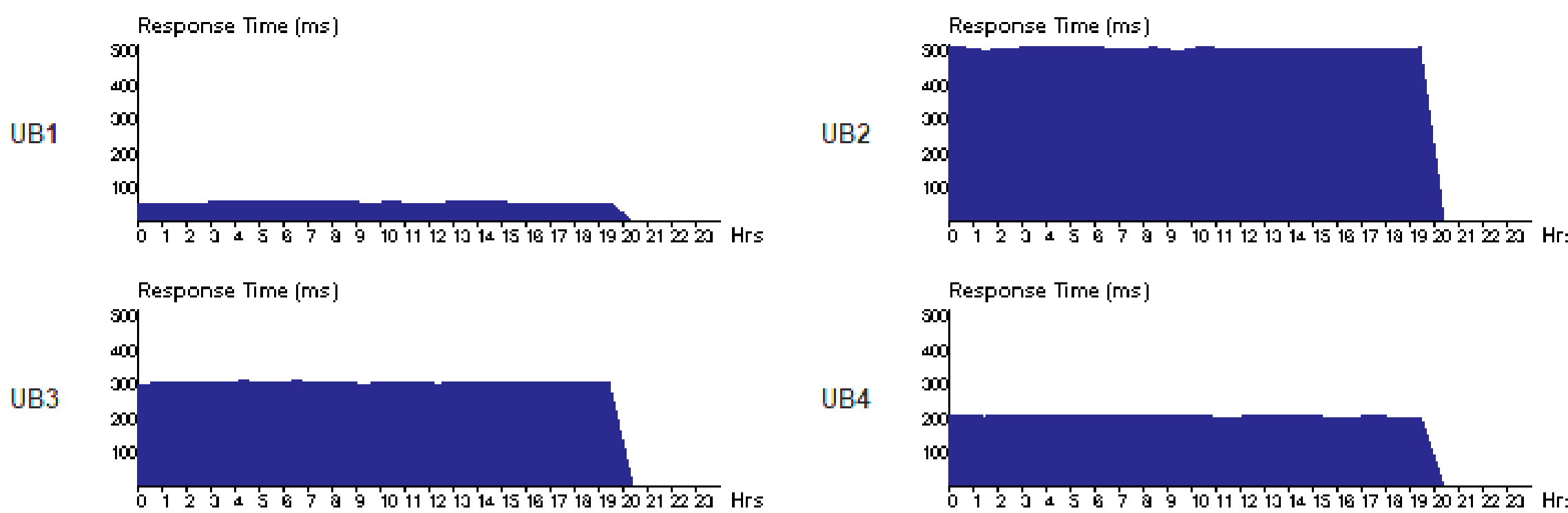

B: Equally Spread

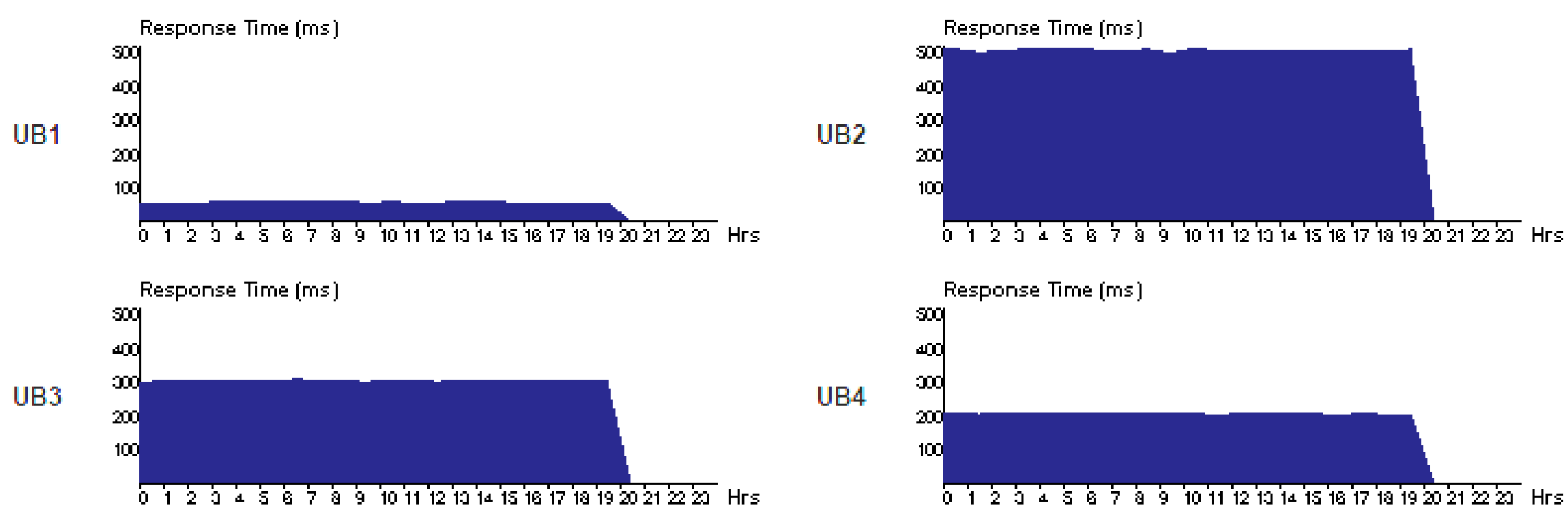

C: Throttled

**Fig. 9** User Base Hourly Response Times with strategy step 2

Figures 7–9 illustrate the temporal behavior of system performance under strategy step 2 for the Round Robin, Equally Spread, and Throttled load balancing strategies across two data centers (DC1 and DC2) and four user bases (UB1–UB4). Figure 7 shows that processing times in both data centers generally follow similar diurnal patterns across strategies, with the Round Robin approach exhibiting lower and more stable processing times, while Equally Spread and Throttled display slightly higher variability, reflecting increased internal workload redistribution. Figure 8

presents the hourly request rates, indicating pronounced demand peaks during early operational hours, after which request volumes decline substantially; this trend is consistent across all strategies and data centers, suggesting that traffic patterns are primarily driven by user demand rather than scheduling policy. Figure 9 depicts hourly response times for individual user bases, where UB1 consistently experiences the lowest response times and UB2 the highest, regardless of the load balancing strategy. While all three strategies demonstrate similar overall response time trends, the Equally Spread and Throttled strategies achieve modest reductions in peak response times for some user bases, indicating improved handling of transient load spikes.

**Table 17:** Result of Cost for Round Robin with strategy step 2

| Data Center | VM Cost | Data Transfer Cost |
|---|---|---|
| OC2 | 100.003 | 8.375 |
| OC1 | 10 | 8.5 |

**Table 18:** Result of Cost for Equally Spread with strategy step 2

| Data Center | VM Cost | Data Transfer Cost |
|---|---|---|
| DC2 | 100.003 | 8.618 |
| DC1 | 100.003 | 8.257 |

**Table 19:** Result of Cost for Throttled with strategy step 2

| Data Center | VM Cost | Data transfer Cost |
|---|---|---|
| DC2 | 100.003 | 8.354 |
| DC1 | 100.003 | 8.521 |

Tables 17–19 present the cost analysis of the Round Robin, Equally Spread, and Throttled load balancing strategies under strategy step 2, considering both virtual machine (VM) costs and data transfer costs across data centers. As shown in Table 17, which reports the results for the Round Robin strategy, DC2 incurs a VM cost of 100.003 and a data transfer cost of 8.375, resulting in a total cost of 108.378. In contrast, DC1 exhibits significantly lower costs, with a VM cost of 10 and a data transfer cost of 8.5, yielding a total cost of 18.5.

The results for the Equally Spread strategy are summarized in Table 18. Both DC2 and DC1 show identical VM costs of 100.003, while their data transfer costs differ slightly, at 8.618 for DC2 and 8.257 for DC1. Consequently, the total costs amount to 108.622 for DC2 and 108.26 for DC1. Table 19 presents the cost outcomes for the Throttled strategy. Similar to the Equally Spread approach, both data centers incur the same VM cost of 100.003. However, data transfer costs are 8.354 for DC2 and 8.521 for DC1, resulting in total costs of 108.358 and 108.524, respectively.

Tables 17–19 illustrate that VM costs dominate total expenditure across all strategies, while variations in data transfer costs introduce modest differences between data centers. The Throttled strategy shows slightly higher data transfer costs compared to Equally Spread, whereas Round Robin demonstrates substantial cost asymmetry between data centers. These results highlight the importance of jointly considering computational and networking costs when selecting cost-effective load balancing strategies for cloud resource management.

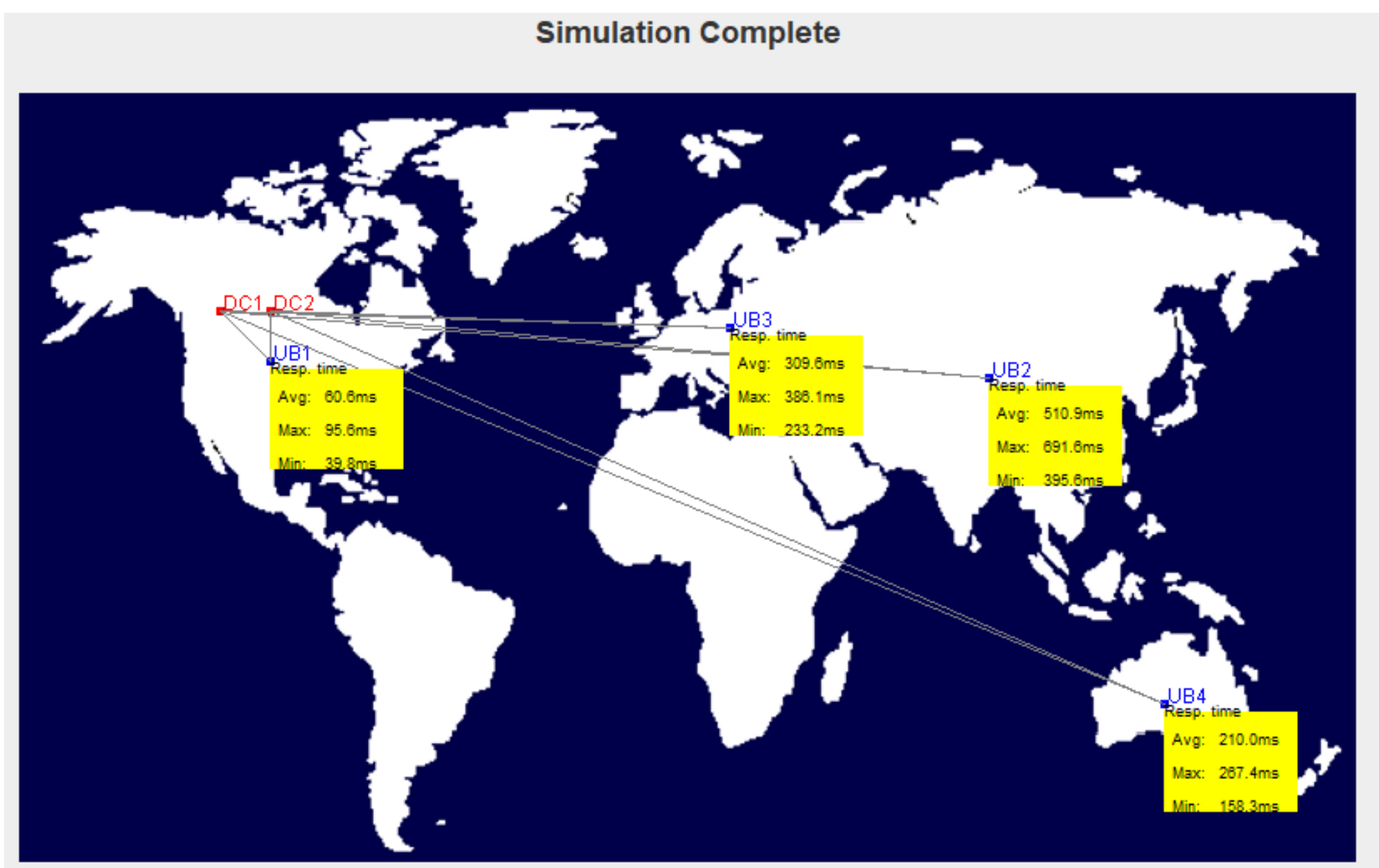

**Fig. 10** Load Balancing technique Round Robin with Strategy step 2

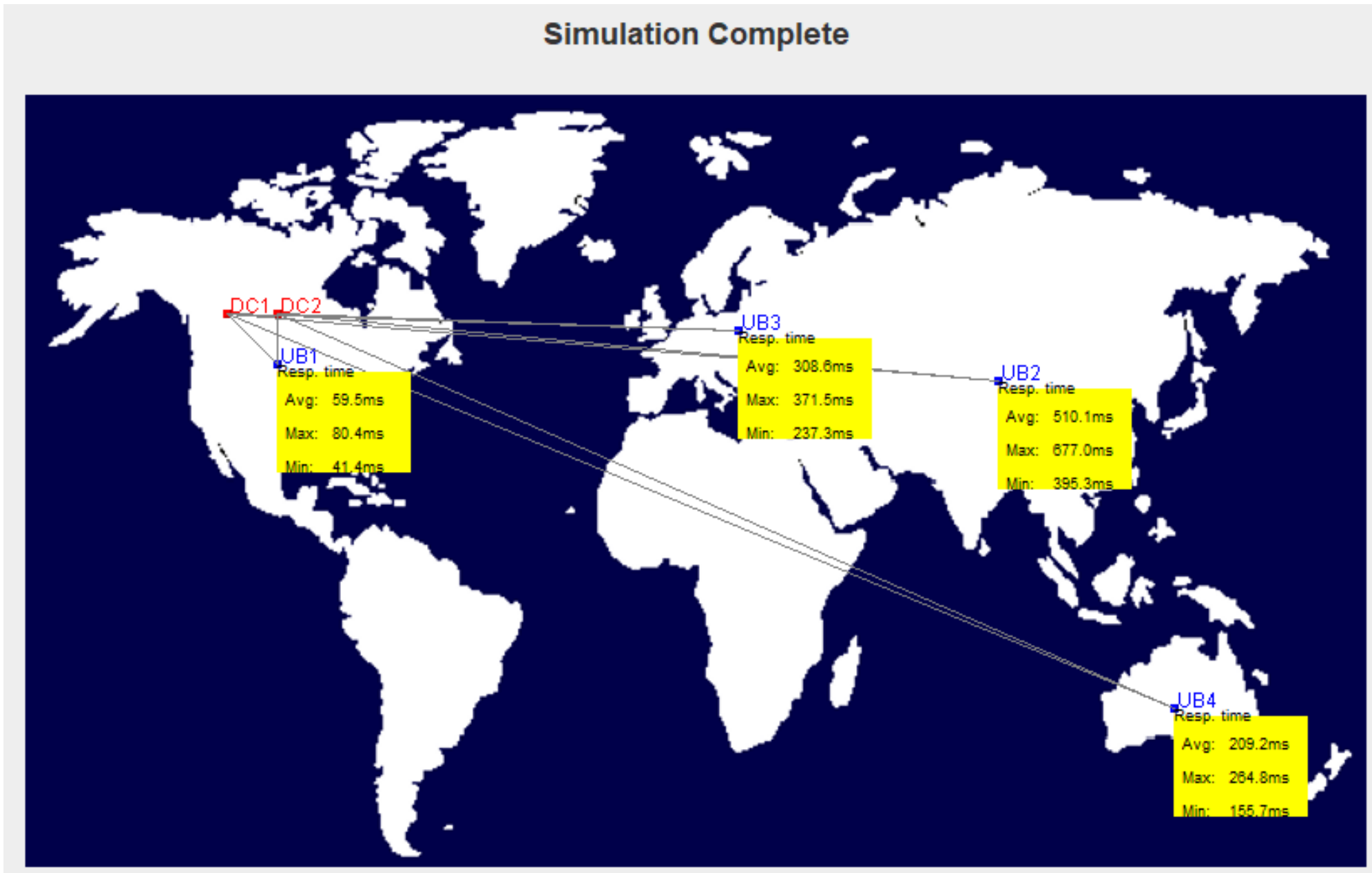

**Fig. 11** Load Balancing technique Equally Spread with Strategy step 2

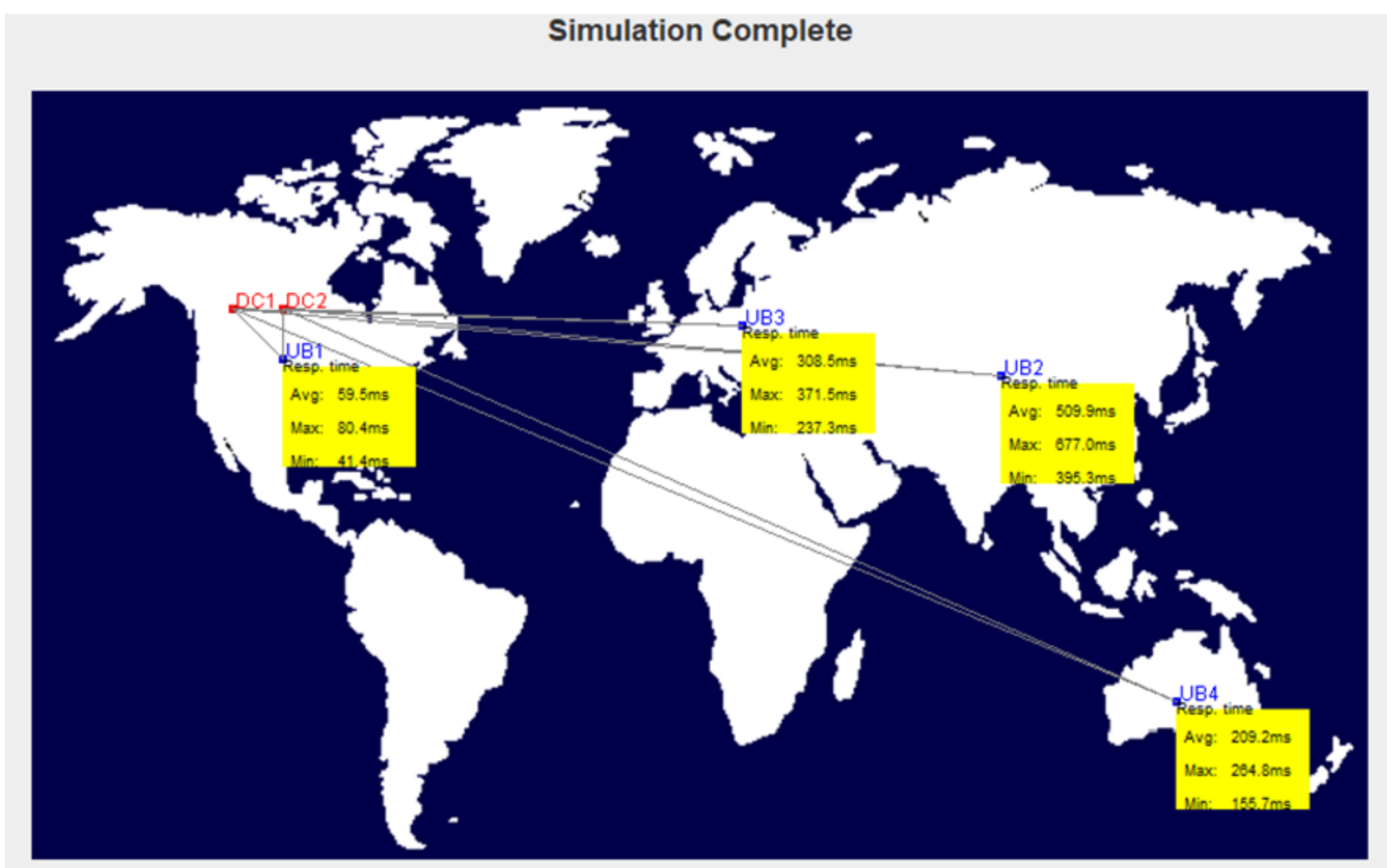

**Fig. 12** Load Balancing technique Throttled with Strategy step 2

According to the average, minimum, and maximum metrics for each user base, the Round Robin technique distributes network requests in a circular order, resulting in varying response times Figure 10). To ensure fairness in handling requests, this strategy is typically used. Equally Spread (Figure 11) distributes network load evenly among all available servers. According to the similar average response times and narrower ranges between minimum and maximum response times, this method may result in more consistent response times across the user bases. Incoming network requests are throttled (Figure 12). By dynamically adjusting the acceptance rate of requests, this strategy prevents server overloads, resulting in more stable server performance and response times. In each image, the overall response time section shows the average, maximum, and minimum response times for each user base. It is crucial to have this information when evaluating a cloud network's performance and choosing the best load balancing strategy. In cloud environments, these simulations assist in visualizing the distribution of response times across data centers and can inform decisions on network configuration and capacity planning.

## 4. Discussion

Various load balancing algorithms across the spectrum of cloud computing are examined and analyzed in the provided studies, highlighting the critical importance and complexity of managing workloads in cloud environments. A common objective of these investigations is to improve the efficiency, reliability, and performance of cloud services while ensuring optimal resource utilization and user satisfaction. It offers valuable insight into the development of more sophisticated load balancing solutions capable of meeting the dynamic demands of cloud computing, ranging from the traditional Round Robin and Throttled algorithms to the more advanced genetic algorithms and hybrid approaches. The need for this sophistication is underscored by the development of highly specialized deep learning models, such as the comparative analysis of U-Net and Segment Anything Model (SAM) for breast tumor detection, which demand robust and responsive computational backbones (Ahmadi et al., 2023). Cloud computing challenges and the industry's efforts to adapt to them are evident in the progression from static to dynamic load balancing algorithms. With their simplicity and predictability, static algorithms provide a foundation for load distribution, but they fall short in environments with variable workloads. Dynamic algorithms, however, allow cloud resources to be managed more efficiently and adaptable through real-time assessments of system states. As the demand for resources fluctuates dramatically and unpredictably in cloud computing, this adaptability is crucial.

In addition, adopting soft computing techniques like genetic algorithms and particle swarm optimization can help load balancers become more flexible and intelligent. In addition to optimizing resource allocation, reducing response times, and improving overall system stability, these methods are inspired by natural processes and behaviors. Based on the comparative analyses and simulations conducted in the studies, bio-inspired algorithms can successfully navigate cloud computing environments despite their complexity. Optimizing cloud computing continues with the enhanced throttled load balancing algorithm and hybrid models. With these advanced models, different algorithms are integrated together and feedback mechanisms and real-time data analysis are incorporated, resulting in better load distribution, reduced resource bottlenecks, and improved scalability and resilience. It is important to note that these innovations reflect not only technical advancements in the field, but also a growing understanding of cloud dynamics and the need for more sophisticated management tools. After collecting the data, we analyze the results from both scenarios. Our objective is to understand:

- How the distribution of resources and data centers affects the overall efficiency of the system.
- What the response times are in each scenario and what differences exist between the two cases.
- How operational costs change in each scenario and which approach is more cost-effective.
- Based on our analysis of the results, we draw conclusions that can aid in improving the design of cloud infrastructures and resource management strategies. Additionally, we provide recommendations for optimizing resource usage in cloud environments, considering the outcomes from the simulations.

### *4.1. Impact of Resource Distribution*

According to the results, intelligent distribution of resources and data centers can significantly improve response times. A shorter response time was observed in the second scenario, where resources were distributed across multiple data centers, compared to the first scenario, where all resources were centralized.

### *4.2. Resource Utilization*

Additionally, data analysis shows that resource distribution improves response times as well as optimizes resource utilization. Distributing the load across several data centers prevents saturation of resources and prevents overuse of any single resource.

### *4.3. Operational Costs*

As a result of setting up and maintaining multiple infrastructures, distributing resources across multiple data centers may incur higher initial costs. This approach, however, can lead to cost reductions in the long run, as improved efficiency and reduced system downtime can lower operating costs. This is particularly relevant for content-delivery applications where optimizing the trade-off between quality and bitrate for various video codecs can lead to significant savings in data transfer and energy costs. (Nia, 2024).

In light of the simulation results, several enhancements are proposed to improve the efficiency of cloud infrastructures. The distribution of resources should be aligned with the application requirements and anticipated traffic loads for intelligent resource distribution. In addition, optimizing data center locations can reduce network latency and enhance response times, possibly by positioning data centers nearer to user bases or incorporating edge computing strategies. Resources must be managed in an agile and flexible manner, using dynamic resource management tools that can adapt to demand changes and maximize resource utilization. In addition, cloud infrastructures need to be evaluated regularly from an economic standpoint in order to ensure that they are optimized. A careful analysis of the return on investment and total cost of ownership for cloud services is required. Technological advancements must be kept up with, particularly serverless computing and containerization. With these technologies, cloud infrastructure costs can be reduced and flexibility can be increased. It is imperative to ensure the security of data and applications by implementing solid security strategies, including data encryption and stringent access controls. The risk of breaches can be significantly mitigated by implementing comprehensive security measures and conducting regular audits. Furthermore, automated monitoring and management tools should be used for constant monitoring of system performance. An autonomously detectable and self-repairable cloud environment is made possible by these tools. As a final point, it is crucial to invest in the training and development of the technical team. As cloud technologies evolve, the team must stay up-to-date and proficient in new tools and practices. By utilizing new technologies effectively, the team will be able to face challenges competently, ultimately leading to optimal cloud infrastructure operation.

## 5. Conclusion

The Cloud Analyst simulation results demonstrate that effective cloud infrastructure planning particularly through optimized data center placement, appropriate load balancing strategies, and efficient resource management, can significantly reduce response times, improve resource utilization, and lower operational costs. The findings highlight that logical and well-distributed cloud resource allocation plays a critical role in enhancing overall system performance, especially in multi–data center environments. The results further indicate that cloud infrastructure design decisions, including the selection of load balancing algorithms, data center locations, and resource management strategies, should be guided by comprehensive performance evaluations rather than static assumptions. By leveraging simulation-based analysis prior to deployment, organizations can identify potential challenges, assess system capabilities, and explore optimization opportunities with reduced risk. This approach enables more informed and confident decision-making regarding cloud infrastructure development and scaling.

In addition, the study emphasizes the importance of evaluating multiple performance metrics—such as response time, processing efficiency, and cost-effectiveness—when assessing load balancing strategies. The results confirm that no single strategy is universally optimal; instead, the effectiveness of each approach depends on workload characteristics and system configuration. As cloud environments continue to grow in scale and complexity, load balancing mechanisms are increasingly transitioning from static to more dynamic and adaptive approaches. Finally, this research underscores the need for continuous evaluation and refinement of cloud infrastructure strategies to maintain efficiency, flexibility, and competitiveness in rapidly evolving technological landscapes. Simulation-based performance assessment provides a valuable foundation for long-term cloud optimization, supporting innovation, scalability, and sustainable growth while enabling organizations to respond effectively to emerging challenges and opportunities.